\documentstyle[11pt,aaspp4,flushrt,tighten]{article}
\singlespace

\def\kmsmpc{\,{\rm km\,s^{-1}\,Mpc^{-1}}}

\def\eblunits{\,{\rm nW\,m^{-2}\,sr^{-1}}}

\def\msun{\,{\rm M_\odot}}
\def\mden{\,{\rm M_\odot\,Mpc^{-3}}}
\def\lden{\,{\rm L_\odot\,Mpc^{-3}}}
\def\sfrd{\,{\rm M_\odot\,yr^{-1}\,Mpc^{-3}}}

\def\page{\vfill\eject}

\def\etal{{et al.\ }}

\def\spose#1{\hbox to 0pt{#1\hss}}
\def\lta{\mathrel{\spose{\lower 3pt\hbox{$\mathchar"218$}}
     \raise 2.0pt\hbox{$\mathchar"13C$}}}
\def\gta{\mathrel{\spose{\lower 3pt\hbox{$\mathchar"218$}}
     \raise 2.0pt\hbox{$\mathchar"13E$}}}

\begin{document}
\title{Deep Galaxy Counts, Extragalactic Background Light, 
and the Stellar Baryon Budget}

\author{Piero Madau\altaffilmark{1,2} and Lucia Pozzetti\altaffilmark{3}}

\altaffiltext{1}{Institute of Astronomy, Madingley Road, 
Cambridge CB3 0HA, UK.}
\altaffiltext{2}{Space Telescope Science 
Institute, 3700 San Martin Drive, Baltimore, MD 21218.}
\altaffiltext{3}{Osservatorio Astronomico di Arcetri, Largo E. Fermi 5,
50125 Firenze, Italy.}
\begin{abstract}

\noindent We assess the constraints imposed by the observed extragalactic 
background light (EBL) on the cosmic history of star formation and the 
stellar mass density today. The logarithmic slope of the galaxy 
number-magnitude relation from the {\it Southern Hubble Deep Field} imaging 
survey is flatter than $0.4$
in all seven $UBVIJHK$ optical bandpasses, i.e. the light from resolved 
galaxies has converged from the UV to the near-IR. We find a lower limit to 
the surface brightness 
of the optical extragalactic sky of about $15\,\eblunits$, comparable to the 
intensity of the far-IR background from {\it COBE} data. 
Assuming a Salpeter initial mass function with a lower cutoff consistent
with observations of M subdwarf disk stars, we set a lower limit of 
$\Omega_{g+s}h^2>0.0013\,I_{50}$
to the visible (processed gas $+$ stars) mass density required to generate an 
EBL at a level of $50\,I_{50}\,\eblunits$; our `best-guess' value is 
$\Omega_{g+s} h^2 \approx 0.0031\,I_{50}$. Motivated by the recent 
microlensing results 
of the MACHO collaboration, we consider the possibility that massive 
dark halos around spiral galaxies are composed of faint white dwarfs, and 
show that only a small fraction ($\lta 5\%$) of the nucleosynthetic baryons 
can be locked in the remnants of intermediate-mass stars forming at $z_F\lta 
5$, as the bright early phases of such halos would otherwise overproduce the 
observed EBL.
  
\end{abstract} 
\keywords{cosmology: miscellaneous -- dark matter -- diffuse radiation -- 
galaxies: evolution -- Galaxy: halo}

\section{Introduction}

Recent progress in our understanding of faint galaxy data made possible 
by the combination of {\it Hubble Space Telescope} {(\it HST}) deep imaging and 
ground-based spectroscopy has vastly improved our understanding of the 
evolution of the stellar birthrate in optically-selected galaxies from the 
present-epoch up to $z\approx 4$ (Steidel \etal 1998; Madau \etal 1998; 
Ellis 1997). The large increase in the quantity of 
information available on the high-redshift 
universe at optical wavelengths has been complemented by measurements 
of the far-IR/sub-mm background by DIRBE and FIRAS onboard the {\it COBE}
satellite (Hauser \etal 1998; Fixsen \etal 1998; Schlegel \etal 1998; 
Puget \etal 1996), showing  
that a significant fraction of the energy released by stellar nucleosynthesis 
is re-emitted as thermal radiation by dust (Dwek \etal 1998), 
and by theoretical progress made in understanding how intergalactic gas follows
the dynamics dictated by dark matter halos until radiative, hydrodynamic, and 
star formation processes take over (Kauffmann \etal 1993; Baugh \etal 1998; 
Somerville \etal 1999; Nagamine \etal 1999). 
Of perhaps equal importance for galaxy formation studies appear the recent 
findings of the microlensing experiments in the direction of the LMC, which
suggest that between 20 and 100\% of the dark matter in the 
Galactic halo is tied up in $0.5^{+0.3}_{-0.2}\,M_\odot$ objects (Alcock \etal 
1997). The underlying 
goal of all these efforts is to understand the growth of cosmic structures,
the internal properties of galaxies and their evolution, and ultimately
to map the star formation history of the universe from the end of
the cosmic `dark age' to the present epoch.

In this paper we focus on the galaxy number-apparent magnitude 
relation and its first moment, the integrated galaxy contribution to the 
extragalactic background light (EBL). The 
logarithmic slope of the differential galaxy counts is a remarkably simple 
cosmological probe of the history of stellar birth in galaxies, as it must 
drop below 0.4 to yield a finite value for the EBL. Together with the 
far-IR/sub-mm background, the optical EBL from both resolved and unresolved 
extragalactic sources is an indicator of the total 
luminosity of the universe, as the cumulative emission from young and evolved 
galactic systems, as well as from active galactic nuclei (AGNs), is recorded 
in this background. As such it provides, for a given stellar mass function, 
a quantitative estimate of the baryonic mass that has been processed by stars 
throughout cosmic history.

Unless otherwise stated, an Einstein-de Sitter (EdS) cosmology ($\Omega_M=1$, 
$\Omega_\Lambda=0$) with $H_0=100\,h\,\kmsmpc$ will be adopted in the 
following. All magnitudes will be given in the AB system.  

\section{Galaxy counts from UV to near-IR}

The HDF-S dataset includes deep near-IR NICMOS images and the deepest 
observations ever made with the STIS 50CCD imaging mode. The new galaxy 
sample we use was extracted from version 1 of the HDF-S catalog created with 
SExtractor (Bertin \& Arnout 1996).  At near-IR wavelengths it consists of 425 
objects detected in the $J+H$ image, over a field of $50^{\prime\prime}\times 
50^{\prime\prime}$. The detection threshold was set to $1 \sigma_{\rm sky}$, 
and the minimum area to 16 drizzled pixels ($0.075^{\prime\prime}$/pxl).
Magnitudes or upper limits in each band were computed from an area 
corresponding to the limiting isophote of the $J+H$ image.
The 50CCD (corresponding roughly to a $V+I$ filter) STIS catalog consists 
of 674 objects detected
again over a field of $50^{\prime\prime}\times 50^{\prime\prime}$ with a 
detection threshold of $0.65 \sigma_{\rm sky}$ over 16 drizzled pixel
($0.025^{\prime\prime}$/pxl). 
Details of the data reduction, source detection algorithm, and photometry are 
can be found on http://www.stsci.edu/ftp/science/hdfsouth/catalogs.html. 

Figure 1 shows the HDF-N and -S galaxy counts compiled directly from the 
catalogs, as a function of AB 
isophotal magnitudes in the $UBVIJHK$ bandpasses for all galaxies 
with signal-to-noise ratio $S/N>3$
within the band. No correction for detection completeness have been made;
in the HDFs the optical counts are likely to be more than $80\%$ 
complete down to this limit (Williams \etal 1996). A compilation of 
existing {\it HST} and ground-based data is also shown. In addition to our 
previous compilation (Pozzetti \etal 1998), we have used $J$ and $H$ 
data from 2MASS (Chester \etal 1998), {\it HST}/NICMOS (Yan \etal 
1998; Teplitz \etal 1998), {\it NTT}/SOFI (Saracco \etal 1999), and 
{\it Keck}/NIRC (Bershady \etal 1998). 
All magnitudes have been corrected to the AB system, while the second order
colour corrections for the differences in the filter effective wavelengths 
have not been applied to the ground-based data.

One should note that different algorithms used for `growing' the
photometry beyond the outer isophotes of galaxies may significantly change
the magnitude of faint galaxies. According to Bernstein \etal (1999), roughly 
50\% of the flux from resolved galaxies with $V>23$ mag lie outside the 
standard-sized apertures used by photometric packages. An extragalactic
sky pedestal created by the overlapping wings of resolved galaxies also 
contributes significantly to the sky level, and is undetectable except 
by absolute surface photometry (Bernstein \etal 1999). Also, at faint 
magnitude levels, distant objects which are brighter than the nominal depth of 
the catalog may be missed due to the $(1+z)^4$ dimming factor.
All these systematic errors are inherent in faint-galaxy photometry;
as a result, our estimates of the integrated fluxes from resolved galaxies 
will typically be too low, and must be strictly considered 
as {\it lower limits}. 

\section{The brightness of the night sky}

The contribution of known galaxies to the optical EBL can be calculated 
directly by integrating the emitted flux times the differential number counts 
down to the detection threshold. The results for $0.35\lta \lambda\lta 
2.2\,\mu$m are listed in Table 1, along with the 
magnitude range of integration and the estimated 1$\sigma$ error bars, which 
arise mostly from field-to-field variations in the numbers of relatively 
bright galaxies.

In all seven bands, the slope of the differential number-magnitude relation is 
flatter than 0.4 above $m_{\rm AB} \sim 20$ (25) at near-IR (optical) 
wavelengths, and this flattening appears to be more pronounced at the shorter 
wavelengths.\footnote{A fluctuation analysis by Pozzetti \etal (1998) has shown
that the turnover observed in the $U$ band is likely due to the `reddening' 
of high redshift galaxies caused by neutral hydrogen along the line of sight.}\,
The leveling off of the counts is clearly seen in Figure 1, where the function 
$i_\nu=10^{-0.4(m_{\rm AB}+48.6)}N(m)$ is plotted against apparent magnitude in
all bands. While counts having a logarithmic slope $d\log N/dm_{\rm AB}=
\alpha\ge0.40$ continue to add to the EBL at the faintest magnitudes, it
appears that the HDF survey has achieved the sensitivity to capture 
the bulk of the near-ultraviolet, optical, and near-IR extragalactic light 
from discrete sources. The flattening at faint 
apparent magnitudes cannot be due to the reddening of distant sources as 
their Lyman break gets redshifted into the blue passband, since the fraction 
of Lyman-break galaxies at (say) $B\approx 25$ is small (Steidel \etal 1996; 
Pozzetti \etal 1998).
Moreover, an absorption-induced loss of sources cannot explain the similar
change of slope of the galaxy counts observed in the $V,I,J,H,$ and $K$ bands.
While this suggests that the surface density of 
optically luminous galaxies is leveling off beyond $z\sim 1.5$, we are 
worried that a significant amount of light may actually be missed at faint 
magnitudes because of systematic errors.

The spectrum of the optical EBL 
is shown in Figure 2, together with the recent results from {\it COBE}.
The value derived by integrating the galaxy counts down to very
faint magnitude levels (because of the flattening of the number-magnitude 
relation most of the contribution to the 
optical EBL comes from relatively bright galaxies) 
implies a lower limit to the EBL intensity in the 0.2--2.2 $\mu$m 
interval of $I_{\rm opt}\approx 15\,\eblunits$. Including the tentative 
detection at 3.5 $\mu$m by Dwek \& Arendt (1999) would boost 
$I_{\rm opt}$ to $\approx 19\,\eblunits$. 
Recent direct measurements of the optical EBL at 3000, 5500, and 8000 \AA\ 
from absolute surface photometry by Bernstein \etal (1999) lie between a 
factor of 2.5 to 3 higher than the integrated light from galaxy counts, with 
an uncertainty that is largely due to systematic rather than 
statistical error. Applying this correction factor to the range 
3000--8000 \AA\ gives a total optical EBL intensity in the range $25-30\,
\eblunits$. This could become $\sim 45\,\eblunits$ if the same correction 
holds also in the near-IR. The {\it COBE}/FIRAS (Fixsen \etal 1998) 
measurements yield $I_{\rm FIR}\approx 14\,\eblunits$ in the 125--2000 $\mu$m 
range. When combined with the DIRBE (Hauser \etal 1998; Schlegel \etal 1998) 
points at 140 and 240 $\mu$m, one gets a far-IR 
background intensity of $I_{\rm FIR}(140-2000\,\mu{\rm m})\approx 
20\,\eblunits$. 
The residual emission in the 3.5 to 140 $\mu$m region is poorly known, but it 
is likely to exceed $10\,\eblunits$ (Dwek \etal 1998). Additional constraints 
-- provided by statistical analyses of the source-subtracted sky -- on the EBL 
have been discussed by, e.g. Martin \& Bowyer (1989), Kashlinsky \etal (1996), 
and Vogeley (1997). 

A `best-guess' estimate 
of the total EBL intensity observed today appears to be
\begin{equation}
I_{\rm EBL}=55\pm 20\,\eblunits. 
\end{equation}
In the following, we will adopt a reference value for the background light 
associated with star formation activity
over the entire history of the universe of $I_{\rm EBL}=50\,I_{50}\eblunits$.  

\section{EBL from quasar activity}

A direct measurement of the contribution of quasars to the EBL 
depends on poorly known quantities like the bolometric correction, 
the faint end of the luminosity function, and the space density of objects 
at high redshifts. Estimates
range from 0.7 to 3 $\eblunits$ (Soltan 1982; Chokshi \& Turner    
1992; Small \& Blandford 1992). Another source of uncertainty is the 
possible existence of a distant population of dusty AGNs with 
strong intrinsic absorption, as invoked in many models for the X-ray 
background (e.g. Madau \etal 1994; Comastri \etal 1995). These Type 
II QSOs, while undetected at 
optical wavelengths, could contribute significantly to the far-IR background.
It is in principle possible to bypass some of the above uncertainties by 
weighing the local mass density of black holes remnants (Soltan 1982), as
recent dynamical evidence indicates that supermassive black holes reside 
at the center of most nearby galaxies (Richstone \etal 1998). The available 
data (about 40 objects) show a correlation between bulge and black hole mass, 
with $M_{\rm BH}\approx 0.006 \,M_{\rm sph}$ as a best-fit (Magorrian \etal 
1998). The mass density in old spheroidal populations today is estimated to be
$\Omega_{\rm sph}h= 0.0018^{+0.0012}_{-0.00085}$ (Fukugita \etal 1998, 
hereafter FHP), implying a mean mass density of quasar remnants today  
\begin{equation}
\rho_{\rm BH}=3\pm 2\times 10^6\,h\, \mden.
\end{equation}
The observed (comoving) energy density 
from all quasars is equal to the emitted energy, $\eta\rho_{\rm BH}c^2$, 
divided by the average quasar redshift, $\langle 1+z\rangle$. Here
$\eta$ is the efficiency of accreted mass to radiation
conversion, equal to 5.7\% for standard disk accretion onto a Schwarzschild
black hole.  The total contribution to the EBL from accretion onto black 
holes can be estimated to be 
\begin{equation}
I_{\rm BH}={c^3\over 4\pi} {\eta \rho_{\rm BH}\over \langle 1+z\rangle}
\approx 4\pm 2.5\, \eblunits\, {\eta\over{0.05}}\, {2.5\over \langle 1+z\rangle}
\end{equation}
($h=0.5$). Therefore, unless dust-obscured accretion onto 
supermassive black holes is a very efficient process ($\eta\gg 0.05$),  
a population of quasars peaking at $z\sim 1.5-2$ is expected to make a 
contribution to the total brightness of the night sky not exceeding 10--20\%
(Fabian \& Iwasawa 1998; Madau 1999).  

\section{The stellar mass density today}

With the help of some simple stellar population synthesis tools we can now 
set a lower limit to the total stellar mass density 
that produced the observed EBL, and constrain 
the cosmic history of star birth in galaxies. One of the most serious 
uncertainties in this calculation is the lower cutoff, usually treated as a 
free parameter, of the initial mass function (IMF).
Observations of M subdwarfs stars with the {\it HST} have recently
shed some light on this issue, showing that the IMF in the Galactic disk can 
be represented analytically
over the mass range $0.1<m<1.6$ (here $m$ is in solar units) by $\log 
\phi(m)={\rm const} -2.33 \log m -1.82(\log m)^2$ (Gould \etal 1996, 
hereafter GBF; Gould \etal 1997). For $m>1$ this 
mass distribution agrees well with a Salpeter function, $\log \phi(m)= {\rm 
const} -2.35 \log m$. A shallow mass function (relative to the Salpeter slope)
below $1\,M_\odot$ has been measured in the Galactic bulge as well (Zoccali 
\etal 1999).  Observations of normal Galactic star-forming regions also show 
some convergence in the basic form of the IMF at intermediate and high masses, 
a power-law slope that is consistent with the Salpeter value (see  Elmegreen 
1998; Massey 1998, and references therein). 
In the following we will use a `universal' IMF (shown in Figure 3) with the 
GBF form for $m<1$, matched to a Salpeter slope for $m\ge 1$; the mass integral of this 
function is 0.6 times that obtained extrapolating a Salpeter function down to 
$0.1\, M_\odot$.\footnote{The bolometric light contributed by stars less 
massive than $1\,M_\odot$ is very small for a `typical' IMF. The use of the 
GBF mass function at low masses instead of Salpeter then leaves the total 
radiated luminosity of a stellar population virtually unaffected.}

As shown in Figure 4, the {\it bolometric} 
luminosity as a function of age $\tau$ of a simple stellar population (a single 
generation of coeval, 
chemically homogeneous stars having total mass $M$, solar metallicity, and the
above IMF) can be well approximated by 
\begin{equation}
L(\tau)= \left\{\begin{array}{ll} 1200\,L_\odot {M\over M_\odot} & 
\mbox{$\tau\le 2.6\,$ Myr;} \\
0.7\,L_\odot {M\over M_\odot} \left({\tau\over 1\,{\rm Gyr}}\right)^{-1.25} 
& \mbox{$2.6\le \tau\le 100\,$ Myr;} \\
2.0\,L_\odot {M\over M_\odot} \left({\tau\over 1\,{\rm Gyr}}\right)^{-0.8} & 
\mbox{$\tau>
100\,$ Myr} \end{array}
\right. \label{eq:bol}
\end{equation}
(cf Buzzoni 1995). 
Over a timescale of 13 Gyr (the age of the universe for an EdS cosmology with 
$h=0.5$), about 1.3 MeV per stellar baryon will be radiated away. This number 
depends only weakly on the assumed metallicity of stars.    
In a stellar system with arbitrary star formation rate per
comoving cosmological volume, $\dot \rho_s$, 
the bolometric emissivity at time $t$ is given by the convolution integral
\begin{equation}
\rho_{\rm bol}(t)=\int_0^t L(\tau)\dot \rho_s(t-\tau)d\tau.
\label{eq:rhob}
\end{equation}
Therefore the total background light observed at Earth ($t=t_H$), 
generated by a stellar population with a formation epoch $t_F$, is 
\begin{equation}
I_{\rm EBL}={c\over 4\pi} \int_{t_F}^{t_H} {\rho_{\rm bol}(t)\over 1+z}dt,
\label{eq:ebl}
\end{equation}
where the factor $(1+z)$ at the denominator is lost to cosmic expansion 
when converting from observed to radiated (comoving) luminosity density. 

To set a lower limit to the present-day mass density, $\Omega_{g+s}$, of 
processed gas $+$ stars 
(in units of the critical density $\rho_{\rm crit}=2.77 
\times 10^{11}\,h^2\mden$), consider now a scenario where all stars 
are formed {\it instantaneously} at redshift $z_F$.
The background light that would be observed at Earth from such an event 
is shown in Figure 4 as a function of $z_F$ for $\Omega_{g+s}h^2=0.0008, 
0.0013, 0.0018$, corresponding to 4, 7, and 9 percent of the 
nucleosynthetic baryon density, $\Omega_bh^2=0.0193\pm 0.0014$ (Burles \& 
Tytler 1998). Two main results are 
worth stressing here: (1) the time evolution of the luminosity radiated by a 
simple stellar population (eq. \ref{eq:bol}) makes the dependence of the 
observed EBL from $z_F$ much shallower than the $(1+z_F)^{-1}$ 
lost to cosmic expansion (eq. \ref{eq:ebl}), as the energy output from
stars is spread over their respective lifetimes; and (2) in order to generate 
an EBL at a level of $50\,I_{50}\,\eblunits$, one requires 
$\Omega_{g+s}h^2>0.0013\,I_{50}$ for an EdS universe with $h=0.5$, hence a mean 
mass-to-blue light ratio today of $\langle M/L_B\rangle_{g+s}>3.5\,I_{50}$ 
(the total blue luminosity density at the present-epoch is ${\cal L}_B=2\times 
10^8\,h\lden$, Ellis \etal 1996). 
The dependence of these estimates on the 
cosmological model (through eq. \ref{eq:ebl}) is rather weak.
With the adopted IMF, about 30\% of this mass will be returned to 
the interstellar medium in $10^8$ yr, after intermediate-mass stars eject 
their envelopes and massive stars explode as supernovae. This return fraction,
$R$, becomes 50\% after about 10 Gyr.\footnote{An asymptotic mass fraction 
of stars returned as gas, $R=\int (m-m_f)\phi(m)dm \times 
[\int m\phi(m)dm]^{-1}\approx 0.5$, can be obtained by using the semiempirical 
initial ($m$)--final ($m_f$) mass relation of Weidemann 
(1987) for stars with $1<m<10$, and by assuming
that stars with $m>10$ return all but a $1.4\msun$ remnant.}

A visible mass density at the level of the above lower limit, while able to 
explain the measured sky brightness, requires (most of the stars) that 
give origin to the observed light to have formed at very low redshifts
($z_F\lta 0.5)$, a scenario which appears to be ruled out by the observed 
evolution of the UV luminosity density (Madau 1999).
For illustrative purposes, it is interesting to consider instead a model where
the star formation rate per unit comoving volume stays approximately constant 
with cosmic time. In an EdS cosmology with $h=0.5$, one derives from equations
(\ref{eq:bol}), (\ref{eq:rhob}), and (\ref{eq:ebl}) 
\begin{equation}
I_{\rm EBL}=1460\,\eblunits \langle {\dot\rho_s\over \sfrd}\rangle.
\end{equation}
The observed EBL therefore implies  a `fiducial' mean star 
formation density of $\langle \dot\rho_s\rangle=0.034$ $I_{50}$ $\sfrd$
(or a factor of 1.6 higher in the case of a Salpeter IMF down to 0.1 
$M_\odot$). Any value much greater than this over a sizeable 
fraction of the Hubble time will generate an EBL intensity well in excess 
of $50\,\eblunits$.
Ignoring for the moment the recycling of returned gas into new stars,
the visible mass density at the present epoch is simply   
$\rho_{g+s}=\int_0^{t_H} \dot \rho_s(t)dt=4.4\times 10^8\,I_{50}\,\mden$, 
corresponding to $\Omega_{g+s}h^2=0.0016\,I_{50}$ 
($\langle M/L_B\rangle_{g+s}=4.4\, I_{50}$). 

Perhaps a more realistic scenario is one where the star formation 
density evolves as 
\begin{equation}
\dot\rho_s(z)={0.23\,e^{3.4z}\over e^{3.8z}+44.7}\,\sfrd. \label{eq:mplot}
\end{equation}
This fits reasonably well all 
measurements of the UV-continuum and H$\alpha$  luminosity densities 
from the present-epoch to $z=4$ after an extinction correction of
$A_{1500}=1.2$ mag ($A_{2800}=0.55$ mag) is applied to the data (Madau 1999), 
and produce a total EBL of about the right magnitude ($I_{50}\approx 1$).  
Since about half of the present-day stars are formed at $z>1.3$ in this model
and their contribution to the EBL is redshifted away,  
the resulting visible mass density is $\Omega_{g+s}h^2=0.0031\,I_{50}$ 
($\langle M/L_B\rangle_{g+s}=8.6\, I_{50}$), almost 
twice as large as in the $\dot \rho_s=$const approximation. 

We conclude that, depending on the star formation history and for the assumed 
IMF, the observed EBL requires between 7\% and 16\% of the nucleosynthetic 
baryon density to be today in the forms of stars, processed gas, and their 
remnants.
According to the most recent census of cosmic baryons, the mass density in 
stars and their remnants observed today is $\Omega_sh=0.00245^{+0.00125} _
{-0.00088}$ (FHP), corresponding to a mean stellar mass-to-blue light 
ratio of $\langle M/L_B\rangle_s=3.4^{+1.7}_{-1.3}$ for $h=0.5$ (roughly 70\% 
of this mass is found in old spheroidal populations). While this is about a 
factor of 2.5 smaller than the visible mass density predicted by equation
(\ref{eq:mplot}), efficient recycling of ejected material into new star 
formation would tend to reduce the apparent discrepancy in the budget. 
Alternatively, the gas returned by stars may be ejected into the intergalactic
medium. With an IMF-averaged yield of returned metals of $y_Z\approx 
1.5\,Z_\odot$,\footnote{Here we have taken $y_Z\equiv \int mp_{\rm zm}\phi(m)dm 
\times [\int m\phi(m)dm]^{-1}$, the stellar yields $p_{\rm zm}$ of Tsujimoto
\etal (1995), and a GBF$+$Salpeter 
IMF.}\, the predicted mean metallicity at the present epoch is 
$y_Z\Omega_{g+s}/\Omega_b\approx 0.25\,Z_\odot$, in good agreement with the values
inferred from cluster abundances (Renzini 1997).

\section{EBL from MACHOs}

One of the most interesting constraints posed by the observed brightness of the
night sky concerns the possibility that a large fraction of the dark mass in 
present-day galaxy halos may be associated with faint white-dwarf (WD) remnants 
of a population of intermediate-mass stars that formed at high redshifts. The 
results of the microlensing MACHO experiment 
towards the LMC indicates that $60\pm 20\%$ of the sought dark 
matter in the halo of the Milky Way may be in the form of 
$0.5^{+0.3}_{-0.2}\,M_\odot$ objects (Alcock \etal 1997). The mass scale is a 
natural one for white dwarfs, a scenario also supported by the lack of a 
numerous spheroidal population of low-mass main sequence stars in the HDF 
(Gould \etal 1998). The total mass of MACHOs inferred within 50 kpc is 
$2^{+1.2}_{-0.7}\times 10^{11}\,M_\odot$, implying a `MACHO-to-blue light' 
ratio for the Milky Way in the range 5 to 25 solar (cf Fields \etal 1998).
If these values were typical of the luminous universe as a whole, i.e. if 
MACHOs could be viewed as a new stellar population having similar 
properties in all disk galaxies, then the cosmological mass density 
of MACHOS today would be $\Omega_{\rm MACHO}=(5-25)\,f_B{\cal L}_B/
\rho_{\rm crit}=(0.0036-0.017)\,f_B\,h^{-1}$, a significant entry in 
the cosmic baryon budget (Fields \etal 1998). Here $f_B\approx 0.5$ is 
the fraction of the blue luminosity density radiated by stellar disks (FHP). 
Note that if MACHOs are halo WDs, the contribution of their progenitors 
to the mass density parameter is several times higher.

Halo IMFs which are very different from that of the solar
neighborhood, i.e. which are heavily-biased towards WD progenitors and 
have very few stars forming with masses below $2\,M_\odot$ (as these would
produce bright WDs in the halo today that are not seen), and above 
$8\,M_\odot$ (to avoid the overproduction of heavy elements), have been 
suggested as a suitable mechanism for explaining the microlensing data
(Adams \& Laughlin 1996; Chabrier \etal 1996). While the halo WD scenario 
may be tightly constrained by the observed rate of Type Ia SN in galaxies 
(Smecker \& Wyse 1991), the expected C and N overenrichment of halo stars 
(Gibson \& Mould 1997), and the number counts of faint galaxies in deep 
optical surveys (Charlot \& Silk 1995), here we explore a potentially more 
direct method (as it is does not depend on, e.g. extrapolating stellar yields 
to primordial metallicities, on galactic winds removing the excess heavy 
elements 
into the intergalactic medium, or on the reddening of distant halos by dust),
namely we will compute the contribution of WD progenitors in dark galaxy halos 
to the extragalactic background light. 

Following Chabrier (1999), we adopt a truncated power-law IMF, 
\begin{equation}
\phi(m)={\rm const}\times e^{-(\overline{m}/m)^3}\, m^{-5}.
\end{equation}    
This form mimics a mass function strongly peaked 
at $0.84\,\overline{m}$. To examine the dependence of the IMF on 
the results we consider two functions (shown in Fig. 3), $\overline{m}=2.4$ and 
$\overline{m}=4$: both yield a present-day Galactic halo mass-to-light ratio 
$>100$ after a Hubble time, as required in the absence of a large non-baryonic 
component. We further assume that a population of halo WD progenitors 
having mass density $X\Omega_bh^2=0.0193X$ formed 
instantaneously at redshift $z_F$ with this IMF and nearly primordial 
($Z=0.02\,Z_\odot$) metallicity. The resulting EBL from such an event is huge, 
as shown in Figure 5 for $X=0.1, 0.3,$ and 0.6 and a $\Lambda$-dominated 
universe with $\Omega_M=0.3$, $\Omega_\Lambda=0.7$, and $h=0.65$ ($t_H=14.5$ 
Gyr).

Consider the $\overline{m}=
2.4$ case first. With $z_F=3$ and $X=0.6$, this scenario would generate 
an EBL at a level of $300\,\eblunits$.
Even if only 30\% of the nucleosynthetic baryons formed at $z_F=5$ with 
a WD-progenitor dominated IMF, the resulting background light at 
Earth would exceed the value of $100\,\eblunits$, the `best-guess' upper limit
to the observed EBL from the data plotted in Figure 2.
The return fraction is $R\approx 0.8$, 
so only 20\% of this stellar mass would be leftover as WDs, the rest 
being returned to the ISM. Therefore, if galaxy halos comprise 100\% 
of the nucleosynthetic baryons, only a small fraction of their mass, 
$X_{\rm WD}\approx 0.2\times 0.30=0.06$ could be in the form of white dwarfs. 
Pushing  the peak of the IMF to more massive stars, $\overline{m}=
4$, helps only marginally. With $\overline{m}=2.4$, the energy radiated 
per stellar baryon over a timescale of 13 Gyr is equal to 2 MeV, 
corresponding to 10 MeV per baryon in WD remnants. A similar value is
obtained in the $\overline{m}=4$ case: because of the shorter lifetimes of 
more massive stars the expected EBL is reduced, but only by 20\% or so 
(see Fig. 5). Moreover, the decreasing fraction of 
leftover WDs raises even more severe problems of metal galactic enrichment.

Note that these limits are not necessarily in contrast with the microlensing
results, as they may imply either that WDs are not ubiquitous in 
galaxy halos (i.e. the Milky Way is atypical), or that the bulk of the baryons 
are actually not galactic. One possible way to relax the above constraints 
on WD baryonic halos is to push their formation epoch to extreme redshifts
($z_F>10$), and hide the ensuing background light in the poorly constrained 
spectral region between  5 and 100 $\mu$m. In Figure 2 we show the EBL 
produced by a WD-progenitor dominated IMF with $\overline{m}=4$ and $(z_F, X, 
X_{\rm WD})=(36, 0.5, 0.1)$, assuming negligible dust reddening. While this 
model may be consistent with the observations if the large corrections factors 
inferred by Bernstein \etal (1999) extend into the near-IR, we draw attention 
to the fact that even a tiny fraction of dust reprocessing in the 
(redshifted) far-IR would inevitably lead to a violaton of the FIRAS background.

\acknowledgments
We have benefited from useful discussions with R. Bernstein, G. Bruzual, 
C. Hogan, A. 
Loeb, and G. Zamorani. We are indebted to R. Bernstein, W. Freedman, \& B.
Madore for communicating their unpublished 
results on the EBL. Partial support for this work was provided by NASA through 
grant AR-06337.10-94A from the Space Telescope Science Institute.

\references

Alcock, C., \etal 1997, ApJ, 486, 697

Armand, C., Milliard, B., \& Deharveng, J.-M. 1994, A\&A, 284, 12 

Baugh, C.~M., Cole, S., Frenk, C. S., \& Lacey, C. G. 1998, ApJ, 498, 504

Bernstein, R. A., Freedman, W. L., \& Madore, B. F. 1999, preprint

Bershady, M. A., Lowenthal, J. D., \& Koo, D. C. 1998, ApJ, 505, 50

Bertin, E., \& Arnouts, S. 1996, A\&AS, 117, 393

Bruzual, A. C., \& Charlot, S. 1993, ApJ, 405, 538  

Burles, S., \& Tytler, D. 1998, ApJ, 499, 699

Buzzoni, A. 1995, ApJS, 98, 69

Chabrier, G. 1999, ApJ, in press (astro-ph/9901145) 

Chabrier, G., Segretain, L., \& Mera, D. 1996, ApJ, 468, L21

Charlot, S., \& Silk, J. 1995, ApJ, 445, 124

Chester, T., Jarret, T., Schneider, S., Skrutskie, M., \& Huchra, J. 1998,
AAS, 192, 5511

Chokshi, A., \& Turner, E. L. 1992, MNRAS, 259, 421

Comastri, A., \etal 1995, A\&A, 296, 1

Dwek, E., \& Arendt, R. G. 1999, ApJ, in press (astro-ph/9809239)

Dwek, E., \etal 1998, ApJ, 508, 106


Elbaz, D., \etal 1998, in The Universe as seen by ISO, ed. P. Cox \& M. F. 
Kessler (Noordwijk: ESA Pub.), in press (astro-ph/9902229) 

Ellis, R.~S. 1997, ARA\&A, 35, 389

Ellis, R.~S., Colless, M., Broadhurst, T., Heyl, J., \& Glazebrook,
K. 1996, MNRAS, 280, 235

Elmegreen, B. G. 1998, in Unsolved Problems in Stellar Evolution, ed. M. Livio
(Cambridge: Cambridge University Press), in press (astro-ph/9811289)

Fabian, A. C., \& Iwasawa, K. 1999, MNRAS, in press

Fields, B.D., Freese, K., \& Graff, D. S. 1998, NewA, 3, 347  

Fixsen, D. J., \etal 1998, ApJ, 508, 123


Fukugita, M., Hogan, C. J., \& Peebles, P. J. E. 1998, ApJ, 503, 518



Gibson, B., \& Mould, J. 1997, ApJ, 482, 98

Gould, A., Bahcall, J. N., \& Flynn, C. 1996, ApJ, 465, 759 (GBF)

Gould, A., Bahcall, J. N., \& Flynn, C. 1997, ApJ, 482, 913

Gould, A., Flynn, C., \& Bahcall, J. N. 1998, ApJ, 503, 798 




Hauser, M. G., \etal 1998, ApJ, 508, 25

Kashlinsky, A., Mather, J. C., Odenwald, S., \& Hauser, M. G. 1996, ApJ, 470,
681

Kauffmann, G., White, S. D. M., \& Guiderdoni, B. 1993, MNRAS, 264, 201

Lagache, G., Abergel, A., Boulanger, F., Desert, F. X., \& Puget, J.-L. 
1999, A\&A, 344, 322 

Madau, P. 1999, in Proceedings of the Nobel Symposium on Particle Physics and 
the Universe, Physica Scripta, in press

Madau, P., Ghisellini, G., \& Fabian, A. C. 1994, MNRAS, 270, L17

Madau, P., Pozzetti, L., \& Dickinson, M. 1998, ApJ, 498, 106 

Magorrian, G., \etal 1998, AJ, 115, 2285

Martin, C., \& Bowyer, S. 1989, ApJ, 338, 677

Massey, P. 1998, in The Stellar Initial Mass Function, ed. G. Gilmore \& D. 
Howell (San Francisco: ASP), p. 17

Nagamine, K., Cen, R., \& Ostriker, J. P. 1999, ApJ, submitted  
(astro-ph/9902372)

Pozzetti, L., Madau, P., Zamorani, G., Ferguson, H. C., \& Bruzual, A. G. 1998, 
MNRAS, 298, 1133

Puget, J.-L., Abergel, A., Bernard, J.-P., Boulanger, F., Burton, W. B., 
Desert, F.-X., \& Hartmann, D. 1996, A\&A, 308, L5
 

Renzini, A. 1997, ApJ, 488, 35 

Richstone, D., \etal 1998, Nature, 395, 14

Saracco, P., D'Odorico, S., Moorwood A., \& Cuby, J. G. 1999, Ap\&SS, 
in press (astro-ph/9904069)

Schlegel, D. J., Finkbeiner, D. P., \& Davis, M. 1998, ApJ, 500, 525


Small, T. D., \& Blandford, R. D. 1992, MNRAS, 259, 725 

Smecker, T. A., \& Wyse, R. 1991, ApJ, 372, 448 

Soltan, A. 1982, MNRAS, 200, 115

Somerville, R. S., Primack, J. R., \& Faber, S. M. 1999, MNRAS, in press
(astro-ph/9802268) 

Steidel, C. C., Adelberger, K., Dickinson, M. E., Giavalisco, M., \& Pettini, 
M. 1998, in The Birth of Galaxies, ed. B. Guiderdoni, F. R. Bouchet, Trinh X. 
Thuan, \& Tran Thanh Van (Gif-sur-Yvette: Edition Frontieres), in press 
(astro-ph/9812167)

Steidel, C.~C., Giavalisco, M., Pettini, M., Dickinson,
M.~E., \& Adelberger, K. 1996, ApJ, 462, L17
 
Teplitz, H. I., Gardner, J. P., Malumuth, E. M., \& Heap, S. R. 1998, ApJ, 507,
L17

Tsujimoto, T., Nomoto, K., Yoshii, Y., Hashimoto, M., Yanagida, S.,
\& Thielemann, F.-K. 1995, MNRAS, 277, 945

Vogeley, M. S. 1997, AAS, 191, 304

Weidemann, V. 1987, A\&A, 188, 74

Williams, R.~E., \etal 1996, AJ, 112, 1335

Yan, L., McCarthy, P. J., Storrie-Lombardi, L. J., \& Weymann, R. J. 1998, ApJ, 503, L19

Zoccali, M., \etal 1999, ApJ, submitted (astro-ph/9906452)

%
%
\page

\begin{deluxetable}{l l l l l}
\small                 
\tablenum{1}
\tablecaption{Integrated Galaxy Light \label{ebl}}
\tablewidth{0 pt}
\tablehead{
\colhead{$\lambda$ (\AA)} & \colhead{AB (range)}& 
\colhead{$\nu I_{\nu}$} & 
\colhead{$\sigma^+$} & \colhead {$\sigma^-$}
}
\startdata
~$3600$     & $18.0$--$28.0$ & $2.87$ & 0.58 & 0.42 \nl
~$4500$     & $15.0$--$29.0$ & $4.57$ & 0.73 & 0.47 \nl
~$6700$     & $15.0$--$30.5$ & $6.74$ & 1.25 & 0.94 \nl
~$8100$     & $12.0$--$29.0$ & $8.04$ & 1.62 & 0.92 \nl
$11000$     & $10.0$--$29.0$ & $9.71$ & 3.00 & 1.90 \nl
$16000$     & $10.0$--$29.0$ & $9.02$ & 2.62 & 1.68 \nl
$22000$     & $12.0$--$25.5$ & $7.92$ & 2.04 & 1.21 \nl
\enddata
\tablecomments{$\nu I_\nu$ is in units of $\eblunits$.}
\end{deluxetable}

\begin{figure}
\vspace{1.5cm}
\plottwo{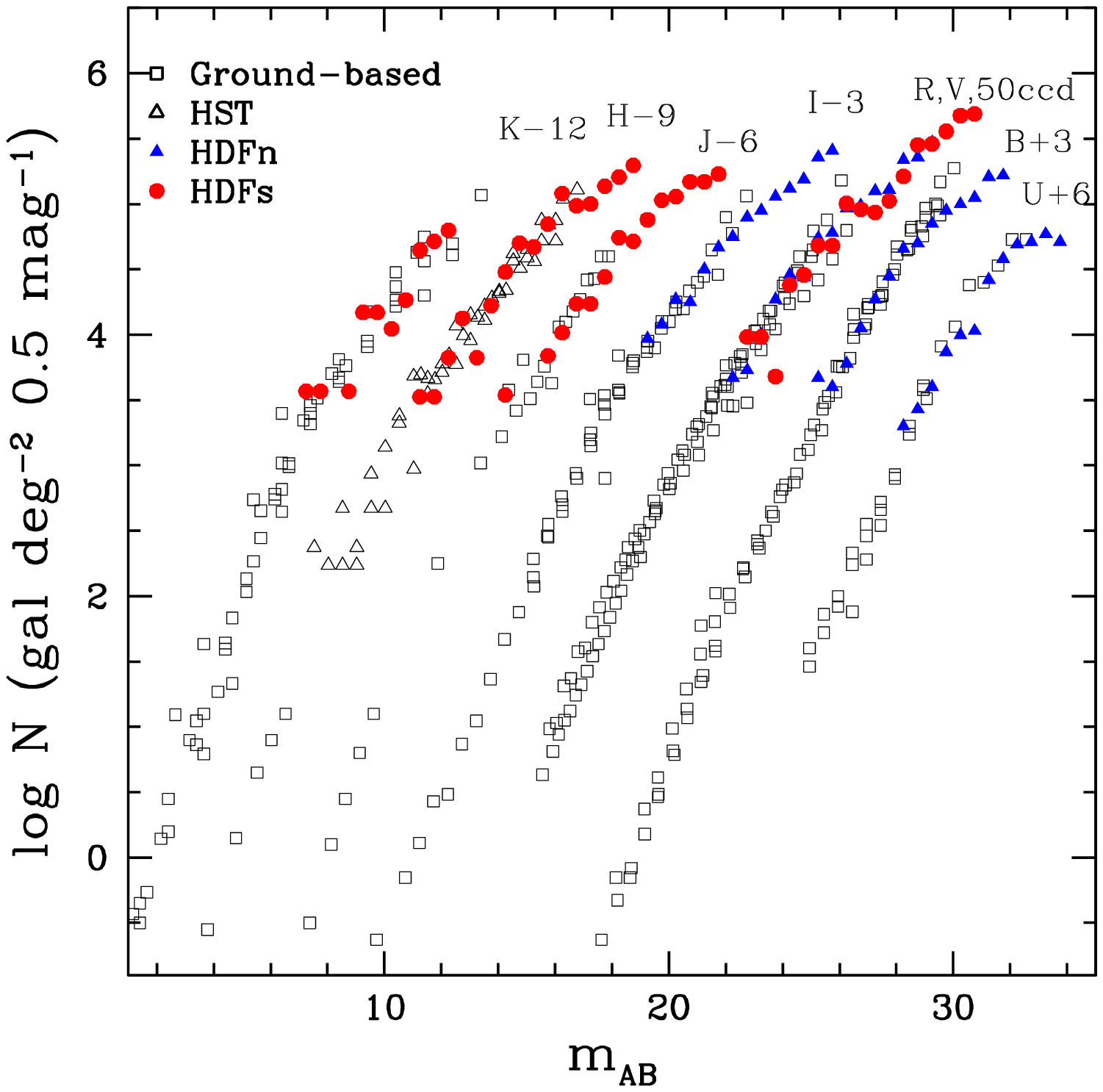}{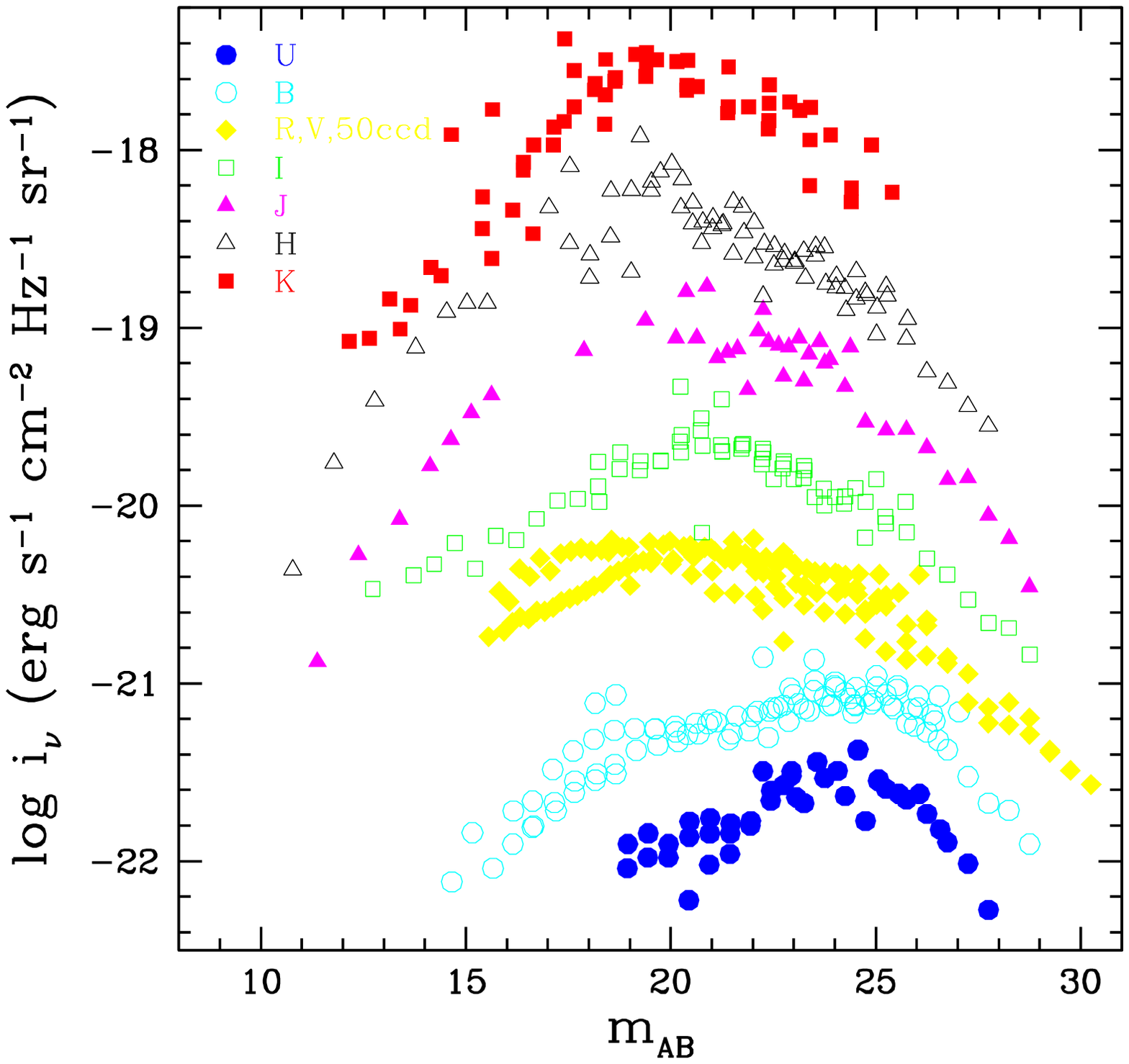}
\caption{\small {\it Left:} Differential $UBVIJHK$ galaxy counts as a function of AB magnitudes.
The sources of the data points are given in the text. Note the decrease
of the logarithmic slope $d\log N/dm$ at faint magnitudes. The flattening is
more pronounced at the shortest wavelengths. {\it Right:}  Extragalactic background 
light per magnitude bin,
$i_\nu=10^{-0.4(m_{\rm AB}+48.6)}N(m)$, as a function of $U$ ({\it
filled circles}), $B$ ({\it open circles}), $V$ ({\it filled pentagons}), $I$
({\it open squares}), $J$ ({\it filled triangles}), $H$ ({\it open triangles}), 
and $K$ ({\it filled squares}) magnitudes. For clarity, the $BVIJHK$ measurements
have been multiplied by a factor of 2, 6, 15, 50, 150, and 600, respectively.
\label{fig1}}
\end{figure}
\vfill\eject

\begin{figure}
\plotone{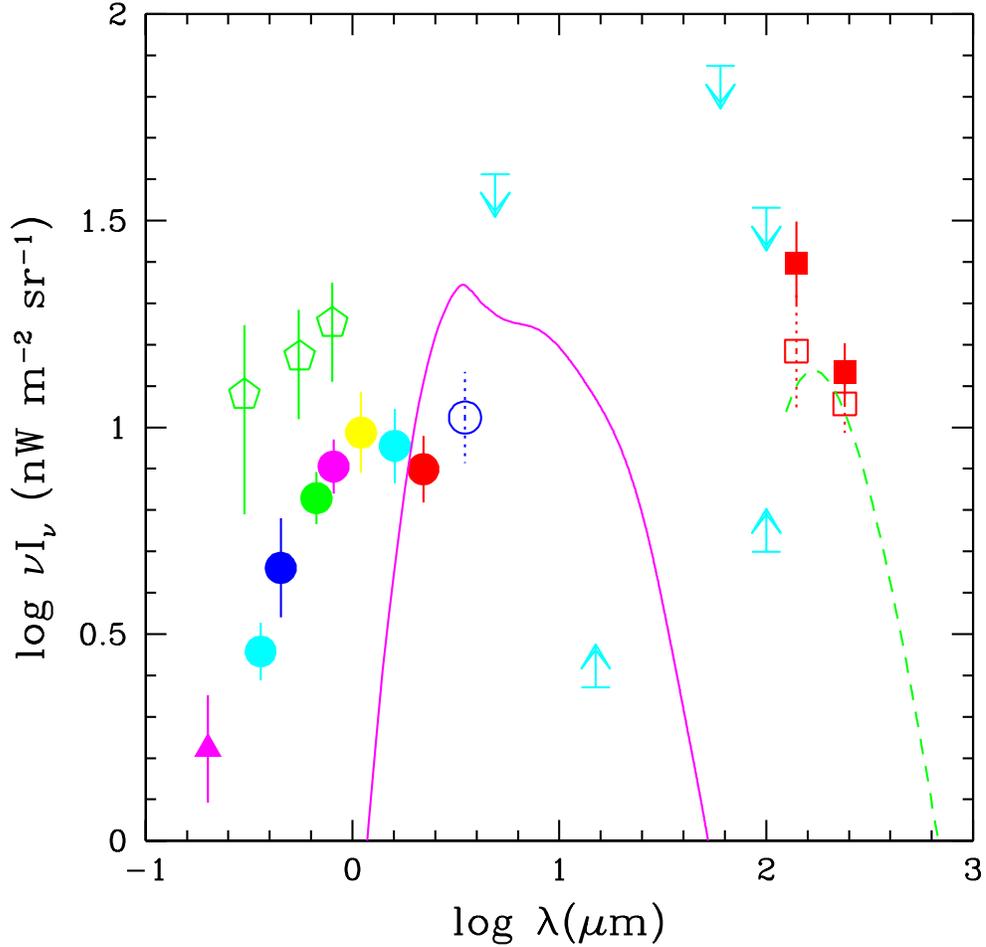}
\vspace{-1.cm}
\caption{\small Spectrum of the optical extragalactic 
background light from resolved sources 
as derived from a compilation of ground-based and space-based galaxy counts 
in the $UBVIJHK$ bands ({\it filled dots}), together with the FIRAS 
125--5000 $\mu$m ({\it dashed line}) and DIRBE 140 and 
240 $\mu$m ({\it filled squares}) detections (Hauser \etal 1998; Fixsen
\etal 1998). The {\it empty squares} show the DIRBE points after correction 
for WIM dust emission (Lagache \etal 1999). Also plotted ({\it filled 
triangle}) is a FOCA-UV point at 2000 \AA\ from Armand \etal (1994), 
and a tentative detection at 3.5 $\mu$m ({\it empty dot}) from 
{\it COBE}/DIRBE observations (Dwek \& Arendt 1999).
The empty pentagons at 3000, 5500, and 8000 \AA\ are Bernstein \etal (1999) 
measurements of the EBL from resolved and unresolved galaxies fainter 
than $V=23$ mag (the error bars showing 2$\sigma$ statistical errors).
Upper limits are from Hauser \etal (1998), the lower limit from Elbaz \etal 
(1999). The {\it solid curve} shows the synthetic EBL produced by 
a WD-progenitor dominated IMF with 
$\overline{m}=4$ and $(z_F, X, X_{\rm WD})=(36, 0.5, 0.1)$, in the case of 
zero dust reddening.
\label{fig2}}
\end{figure}

\begin{figure}
\plotone{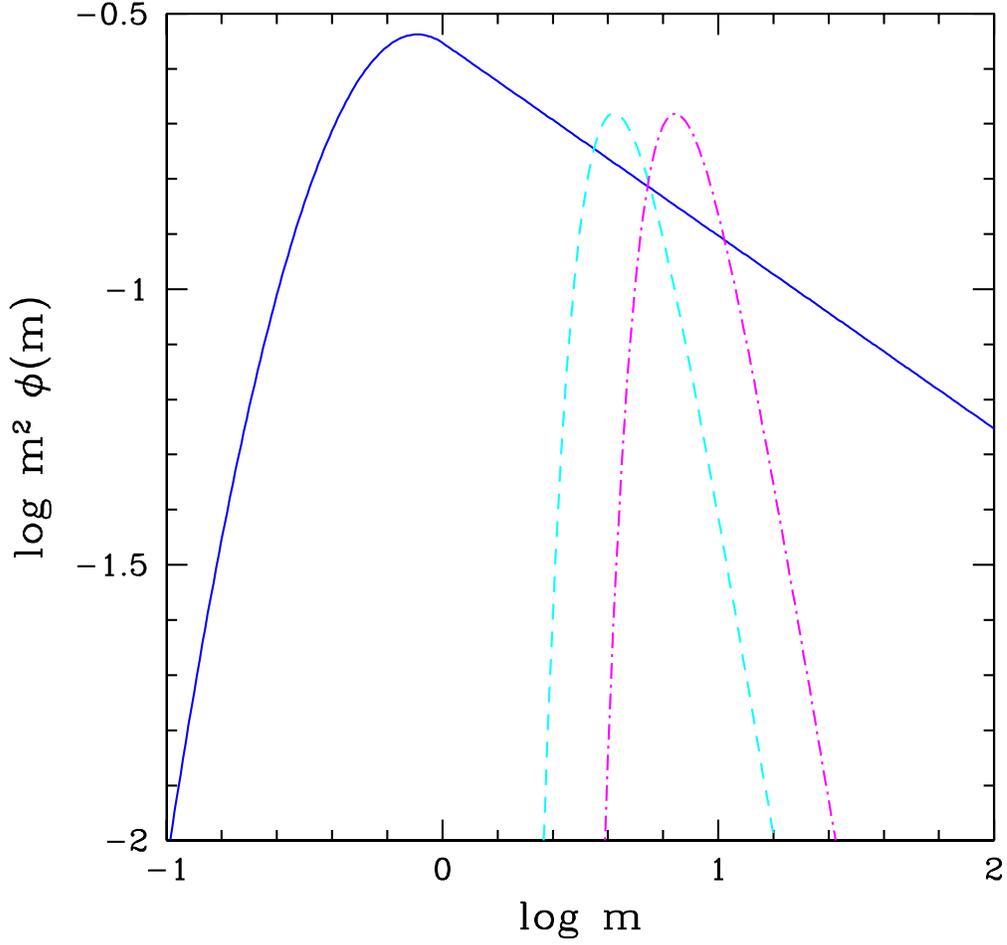}
\caption{\small Stellar initial mass functions, $\phi(m)$, multiplied by $m^2$. 
{\it Solid line:} Salpeter IMF, $\phi(m)\propto m^{-2.35}$ 
at high masses, matched to a GBF function at $m\le 1$. {\it Dotted line:} 
WD-progenitor dominated IMF in galaxy halos, $\phi(m)\propto 
e^{-(\overline{m}/ 
m)^3}\, m^{-5}$, with $\overline{m}=2.4$ (see text for details). 
{\it Dot-dashed line:} Same for $\overline{m}=4$.
All IMFs have been normalized to $\int m\phi(m)dm=1$.
\label{fig3}}
\end{figure}

\begin{figure}
\plottwo{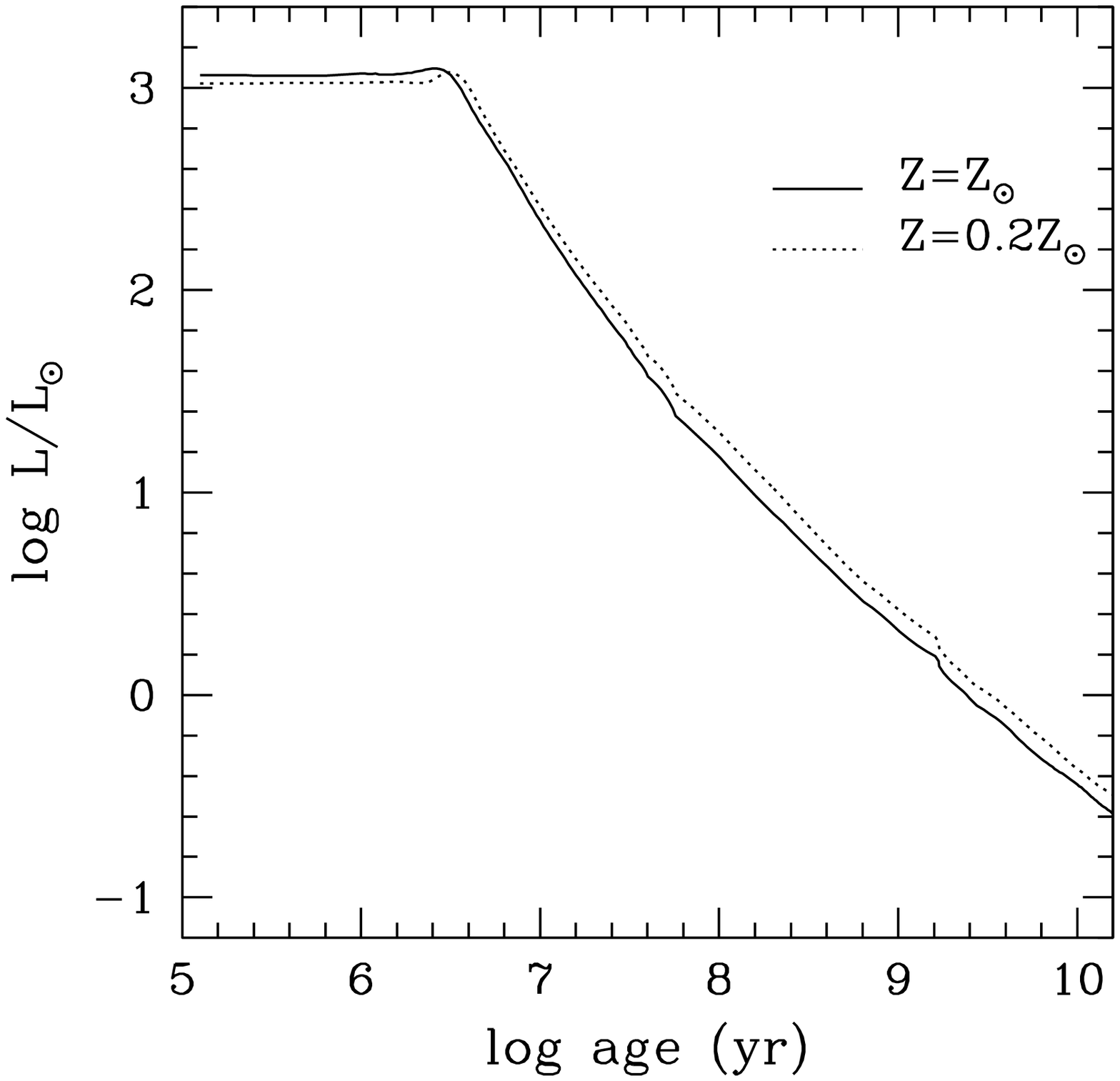}{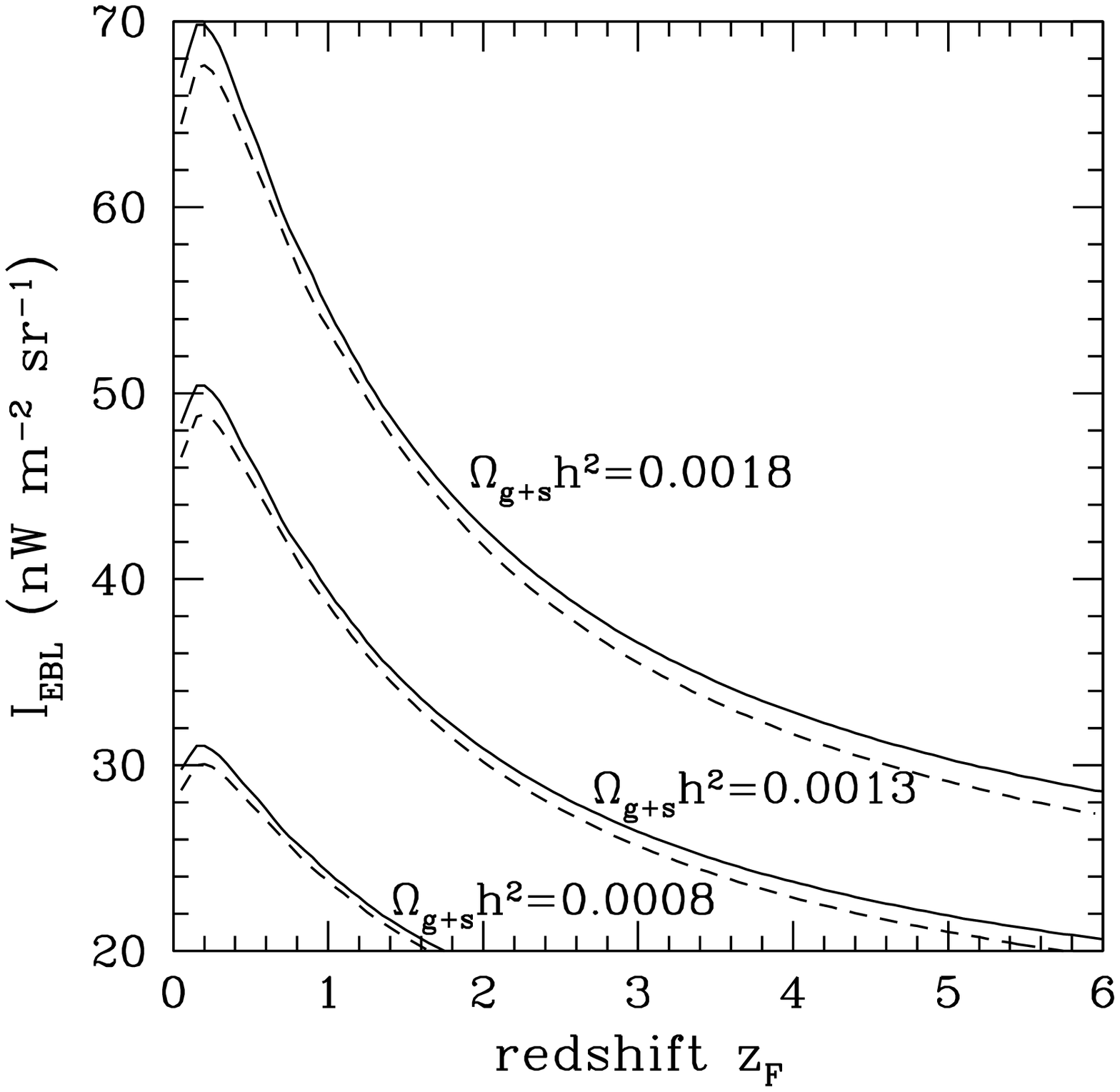}
\caption{\small {\it Left:} Synthetic (based on an update of Bruzual \& Charlot's 1993 
libraries)
bolometric luminosity versus age of a simple stellar 
population having total mass $M=1\,M_\odot$, metallicity $Z=Z_\odot$
({\it solid line}) and $Z=0.2\,Z_\odot$ ({\it dotted line}), and 
a GBF$+$Salpeter IMF (see text for details). 
{\it Right:} EBL observed at Earth from the instantaneous formation at 
redshift $z_F$ of a stellar population having the same IMF ($Z=Z_\odot$)
and mass density $\Omega_{g+s}h^2=0.0018, 0.0013,$ and 0.0008, 
as a function of  
$z_F$. {\it Solid curves:} EdS universe with $h=0.5$ ($t_H=13$ Gyr). 
{\it Dashed curves:}  
$\Lambda$-dominated universe with $\Omega_M=0.3$, $\Omega_\Lambda=0.7$, and
$h=0.65$ ($t_H=14.5$ Gyr).
\label{fig4}}
\end{figure}

\begin{figure}
\plottwo{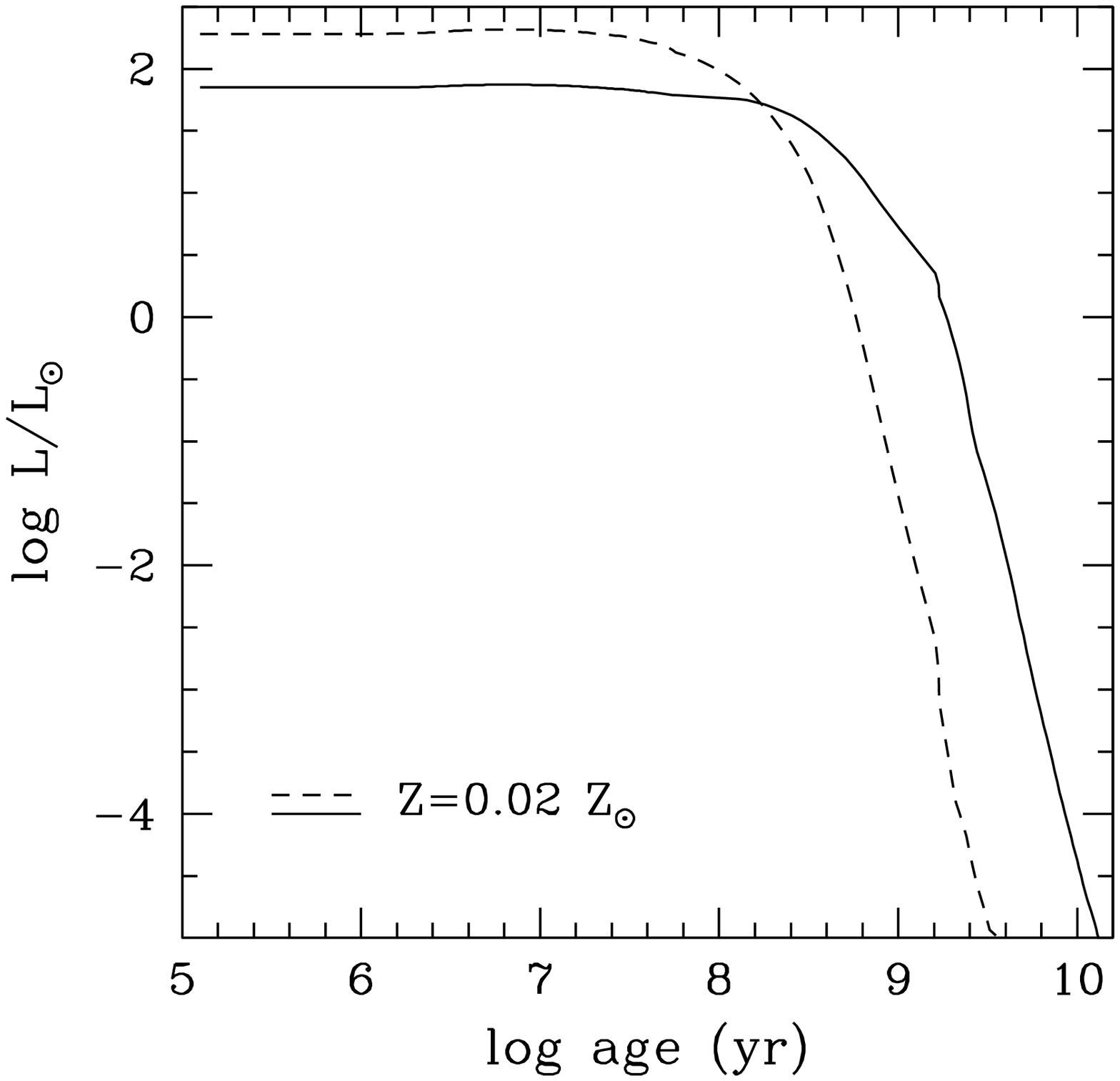}{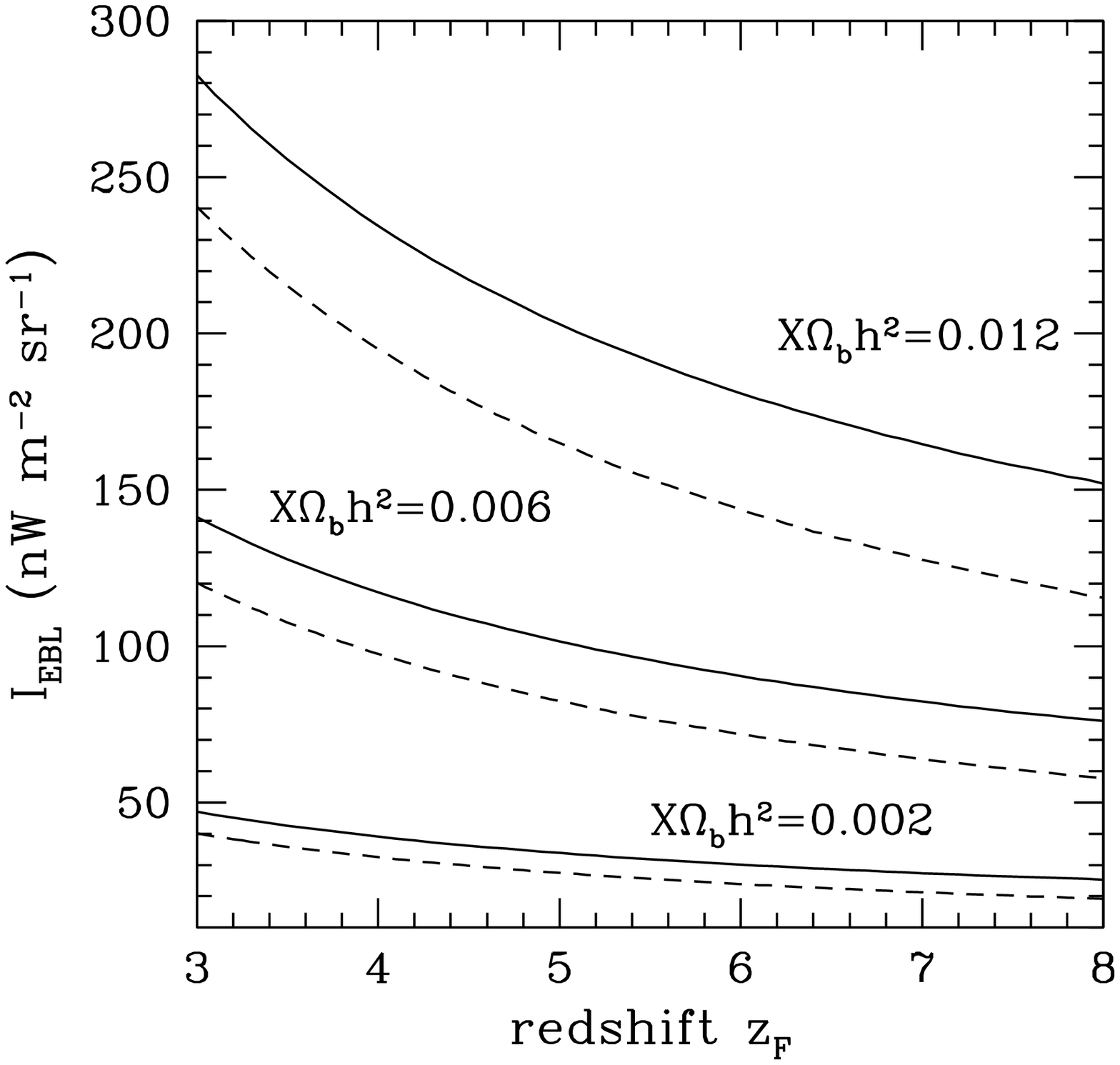}
\caption{\small {\it Left:} Synthetic bolometric luminosity versus age of a simple 
stellar population having total mass $M=1\,M_\odot$, metallicity $Z=0.02
Z_\odot$, and a WD-progenitor dominated IMF (see text for details) 
with $\overline{m}=2.4$
({\it solid line}) and $\overline{m}=4$ ({\it dashed line}). 
{\it Right:} EBL observed at Earth from the instantaneous formation at 
redshift $z_F$ of a stellar population having the same IMF and metallicity, and 
mass density $X\Omega_bh^2=0.012, 0.006$ and 0.002 (corresponding to 60, 30, 
and 10 per cent of the nucleosynthetic value), as a function of  
$z_F$. A $\Lambda$-dominated universe with $\Omega_M=0.3$, $\Omega_\Lambda=
0.7$, and $h=0.65$ has been assumed. {\it Solid line}:
$\overline{m}=2.4$. {\it Dashed line}: $\overline{m}=4$. 
\label{fig5}}
\end{figure}

\end{document}